# A Novel Behavior-Based Recommendation Methodology for E-commerce

Reza Barzegar Nozari [1,*], Mahdi Divsalar [1], Sepehr Akbarzadeh Abkenar [1], Mohammadreza Fadavi Amiri [1], Ali Divsalar [2]

barzegar@shomal.ac.ir, divsalarmahdi@shomal.ac.ir, akbarzadehsepehr@shomal.ac.ir, fadavi@shomal.ac.ir, ali.divsalar@nit.ac.ir

[1] Computer Engineering Department, Shomal University, Amol, Iran

[2] Department of Industrial Engineering, Babol Noshirvani University of Technology, Babol, Iran

* Corresponding Author

**Abstract**

The majority of existing recommender systems rely on user ratings, which are limited by the lack of user collaboration and the sparsity problem. To address these issues, this study proposes a behavior-based recommender system that leverages customers' natural behaviors, such as browsing and clicking, on e-commerce platforms.

The proposed recommendation methodology involves clustering active customers, determining neighborhoods, collecting similar users, calculating product reputation based on similar users, and recommending high-reputation products. To overcome the complexity of customer behaviors and traditional clustering methods, an unsupervised clustering approach based on product categories is developed to enhance the recommendation methodology.

This study makes notable contributions in several aspects. Firstly, a groundbreaking behavior-based recommendation methodology is developed, incorporating customer behavior to generate accurate and tailored recommendations leading to improved customer satisfaction and engagement. Secondly, an original unsupervised clustering method, focusing on product categories, enables more precise clustering and facilitates accurate recommendations. Finally, an approach to determine neighborhoods for active customers within clusters is established, ensuring grouping of customers with similar behavioral patterns to enhance recommendation accuracy and relevance.

Overall, this research presents an innovative procedure for enhancing recommendations on e-commerce platforms through behavioral data. The proposed recommendation methodology and clustering method contribute to improved recommendation performance, offering valuable insights for researchers and practitioners in the field of e-commerce recommendation systems. Additionally, the proposed method outperforms benchmark methods in experiments conducted using a behavior dataset from the well-known e-commerce site Alibaba.

## 1. Introduction

The rapid expansion of e-commerce platforms has revolutionized people's daily shopping habits and behaviors. Today, millions of individuals rely on online platforms to fulfill their shopping needs. On the other hand, these platforms offer an extensive array of products and services, creating a highly competitive environment for e-shops to thrive and for customers to find their desired items among a vast selection (Roy and Ludwig 2021). Recognizing the need to earn customer loyalty over time, e-shops strive to provide a seamless experience, while customers desire to find suitable products efficiently. Recommender Systems (RS) have emerged as a solution, benefiting both e-shops and customers. RS assists customers in discovering products or services aligned with their interests, consequently enhancing customer loyalty (Liao and Chang 2016). In other words, e-shops employ RS to suggest suitable products and cultivate customer loyalty. Therefore, many leading e-commerce platforms such as Alibaba, Amazon, and eBay have implemented RS to excel in this competitive landscape (Schafer et al. 2001; Zhao et al. 2015).

While various types of RS exist, they can generally be classified into three categories: Content-based, Collaborative Filtering, and Hybrid (Bobadilla et al. 2013). Content-based RS techniques utilize product information to identify similar items based on customers' preferences and recommend them. Collaborative filtering techniques leverage customers' preferences or similar users to identify suitable products and make recommendations. Hybrid RS techniques combine content-based and collaborative filtering methods to find or recommend suitable products. Among these approaches, collaborative filtering techniques have demonstrated remarkable success and widespread utilization, surpassing other types of RS (Koohi and Kiani 2017). Consequently, the proposed recommender system in this study is developed based on the collaborative filtering framework. Furthermore, content-based techniques are not ideal for recommending products that lack easily analyzable attribute information based on customers' past preferences (Kim et al. 2005). Additionally, hybrid methods that combine collaborative filtering and content-based techniques may encounter computation and time complexity issues. Therefore, collaborative filtering is considered the most suitable framework for the proposed methodology in this study.

However, most collaborative filtering methods primarily rely on users' rating information and may not effectively leverage behavioral information, such as viewing, adding to cart, and purchasing, which are prevalent in e-commerce data (Kim et al. 2005). Moreover, rating-based collaborative filtering approaches depend on customers' cooperation to provide ratings for purchased products. Due to the lack of customer participation in rating products, a significant portion of products lacks ratings, resulting in data sparsity issues. As a result, collaborative filtering techniques struggle to recommend unrated products and face challenges in identifying similar customers (Koohi and Kiani 2020). To address these limitations, the proposed method in this study is developed based on customers' behaviors, specifically their browsing and clicking activities on the platform. This approach introduces a new collaborative filtering technique that effectively utilizes behavioral information and resolves the shortcomings of rating-based methods. Customer behavior can be captured through internet technology, enabling the recording of online actions (referred to as click-stream data). Further details on collaborative filtering, behavior analysis, and related work are extensively discussed in the literature.

This article presents an innovative approach to enhancing recommendations on e-commerce platforms through a novel unsupervised clustering method and recommendation methodology based on customers' behaviors. The proposed behavior-based recommender system merges the clustering and recommendation methods, resulting in improved recommendation performance compared to state-of-the-art approaches. The entire process of the proposed behavior-based recommendation methodology is described in detail in the methodology section. In summary, this study makes the following notable contributions:

1. Novel Behavior-Based Recommender System: A groundbreaking recommender system is developed specifically for e-shops, leveraging customer behavior as a key factor in generating recommendations. This innovative approach takes into account the unique browsing, clicking, and purchasing patterns of customers to provide more accurate and tailored recommendations.

2. New Recommendation Methodology: A fresh recommendation methodology is designed, highlighting the importance of customer behavior. By analyzing and understanding customer actions, this methodology enhances the precision and effectiveness of recommendations, ultimately leading to improved customer satisfaction and engagement.

3. Introduction of Unsupervised Clustering Method: An original unsupervised clustering method is introduced, with a particular focus on product categories. By categorizing customers based on their behavior within specific product subgroups, this method enables more precise clustering and subsequently facilitates more accurate recommendations.

4. Neighborhood Determination Approach: An approach is established to determine neighborhoods for active customers within clusters, taking into account the length of their behavioral data. This approach ensures that customers with similar behavioral patterns are grouped together, enhancing the accuracy of recommendations and fostering a sense of relevance for the customers.

The remainder of this article is organized as follows: Section 2 provides a review of the relevant literature. Section 3 explains the proposed method in detail. Section 4 presents the experimental setup and results. Section 5 offers a discussion of the findings. Finally, Section 6 provides the conclusion and outlines future directions for research.

## 2. Literature

This section provides the necessary background for this study by reviewing collaborative filtering techniques (subsection 2.1), behavior analysis (subsection 2.2), and related works (subsection 2.3). It aims to broaden the understanding of collaborative filtering methods and behavior analysis and establish a foundation for the proposed novel approach.

### 2.1. Collaborative Filtering

This subsection delves into collaborative filtering techniques, offering a comprehensive understanding of their principles and applications. It serves to broaden the knowledge base related to collaborative filtering methods.

Collaborative Filtering (CF) stands as one of the most widely used techniques in RSs. Its foundation lies in the theory that people tend to be interested in things that their friends or similar individuals like. Thus, CF techniques aim to recommend new products or predict their suitability for a specific customer by analyzing the interests and preferences of similar customers. The general procedure of CF is depicted in Figure 1. CF techniques offer several advantages over other RS techniques, with the primary benefit being the improved performance of RSs (Ambulgekar et al. 2019).

CF strategies can be classified into two main categories: Memory-based and Model-based. The memory-based method considers all customers and their entire history of preferences to make predictions and recommendations. On the other hand, the model-based method employs machine learning techniques, such as clustering (Koohi and Kiani 2016), neural networks (Nazari et al. 2022), and swarm intelligence (Bedi and Sharma 2012; Pirozmand et al. 2021), to learn an efficient model from customers' past preferences. This learned model is then used for prediction and recommendation.

The core of CF technique lies in the concept of similarity measures, such as Cosine similarity, Pearson correlation coefficient, and Jaccard index, to identify customers who have similar preferences to an active customer (Ambulgekar et al. 2019). Typically, these similarities are computed based on the ratings or explicit feedback provided by users/customers. However, this approach encounters challenges such as the cold start problem and data sparsity when customers do not actively rate products. To address these issues, researchers have proposed incorporating trust statements into the CF process (R. Barzegar Nozari et al. 2020). Trust statements can be either explicit, where individuals directly provide trust ratings about others, or implicit, where trust statements are inferred from individuals' preferences or explicit trust ratings (Reza Barzegar Nozari and Koohi 2021, 2022). Matrix Factorization (Yu et al. 2013) and Deep Representation (Fan et al. 2019) techniques have also been employed in modern CF methods to tackle these challenges and enhance performance.

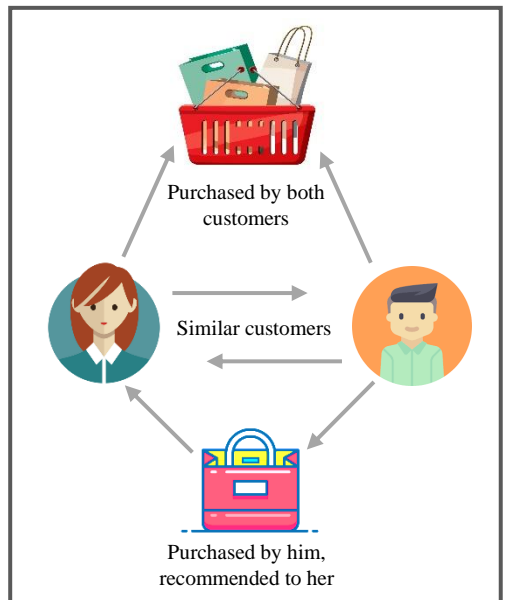

**Figure 1.** General procedure of CF approaches.

In terms of recommendation, most CF methods predict the suitability or rating of products and subsequently recommend one or multiple products with high rankings or ratings. One of the well-known prediction methods in CF is the Resnick function (Reza Barzegar Nozari and Koohi 2021), which heavily relies on the correlation weights between active customers and similar ones. Different articles have proposed various correlation weights, including similarity, trust, confidence, opinion, and learning weights, as alternative formulations of the Resnick function (R. Barzegar Nozari et al. 2020; Reza Barzegar Nozari and Koohi 2021; Koohi and Kiani 2016).

The aforementioned explanation highlights the popularity of CF among researchers for investigation and study purposes. CF-based RS models have found wide applications across domains such as e-commerce, social networks, media, healthcare, and tourism in the literature (Ambulgekar et al. 2019). Additionally, CF techniques are utilized in group recommendation systems (Reza Barzegar Nozari and Koohi 2020), which pose unique challenges in decision-making by recommending activities to a group of individuals (e.g., family or friends) who intend to participate together, such as watching a movie or planning a vacation. To maintain focus within the scope of this study, the provided information about CF is considered sufficient. Further details on CF can be explored in (Ambulgekar et al. 2019).

## 2.2. Behavior Analysis

Focusing on behavior analysis, this subsection explores the significance of studying customer actions such as browsing, clicking, and purchasing behaviors. It highlights the importance of incorporating behavior-based approaches in recommender systems.

Behaviors reflect the actions individuals take to achieve their objectives. In the context of e-commerce, customers' behaviors encompass actions such as product viewing, adding items to favorites, and making purchases (Alfian et al. 2019; Schafer et al. 2001). Behavior analysis plays a vital role in helping marketers understand how customers respond to new products or services, as it has a positive impact on profitability and sales. Consequently, numerous marketing approaches have been developed to gain insights into customer behavior (Chiang et al. 2013; DiClemente and Hantula 2003; T. C. K. Huang 2012; Su and Chen 2015). To conduct behavior analysis, e-commerce platforms need to capture and analyze the online actions of customers.

Thanks to advancements in internet technology, it is now feasible to accurately record customers' online actions and gather a wealth of information (C.-H. Lee et al. 2021). Figure 2 illustrates the various activities and stages that consumers can undertake on an e-commerce platform, ranging from logging in to making a purchase, along with the information that can be derived from these actions. The recorded customer actions, known as click-stream data (Su and Chen 2015), represent a valuable resource for researchers aiming to understand customer behaviors and choices.

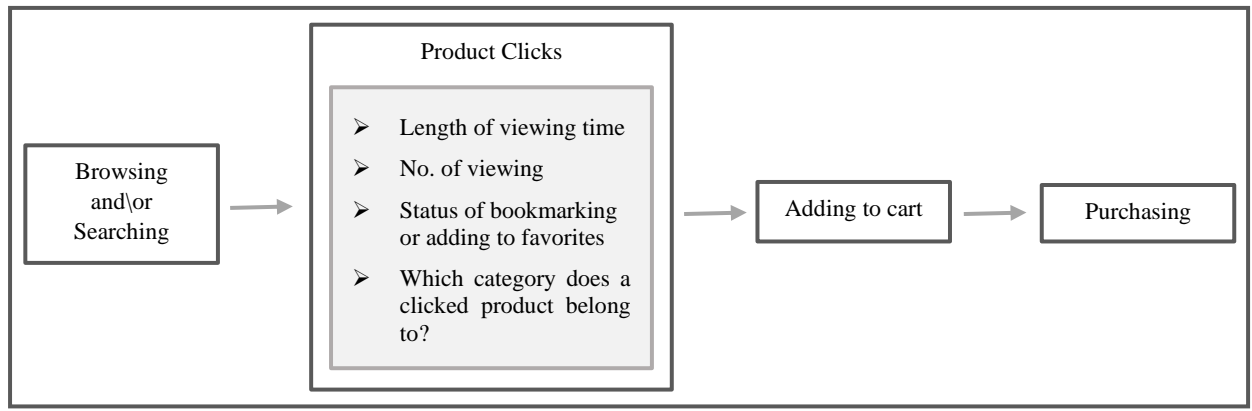

**Figure 2.** Possible actions that customers perform in e-commerce platforms and possible information that can be brought from such actions.

The initial investigations of customer behavior by scholars were inspired by their background in behavioral research (Liao and Chang 2016). These studies emphasized the role of behavior analysis and its application in understanding customer behavior. As interest in the behavioral-based approach grew, researchers focused on customers' perspectives and behaviors, particularly in social marketing and pro-social platforms (DiClemente and Hantula 2003). The success of these studies prompted researchers to employ behavior analysis in the realm of e-commerce. Subsequent research efforts explored different modeling methods to comprehend customers' behavior, particularly in the context of product purchases. In recent years, scholars have utilized diverse techniques such as fuzzy sets (Alrezaamiri et al. 2019; T. C. K. Huang 2012), machine learning (Adeniyi et al. 2014), and association rules (Liao and Chang 2016) to investigate effective strategies for recommending products based on customers' behaviors. The findings of these studies demonstrate that incorporating behavior analysis improves the quality of recommendations. The following subsection will discuss several related works in this field.

### 2.3. Related Works

In this subsection, a review of relevant studies closely aligned with the present research is presented. This review provides insights into previous works that share similarities with the current study, establishing a foundation for the proposed novel approach.

The rise of online shopping applications has led to a surge in behavior analytics within e-commerce, with researchers using customers' purchase patterns, website visit data, and browsing paths to develop models for predicting their preferences (Chiang et al. 2013). To measure customers' interest, various aspects of their behavior are analyzed, including product ratings, purchasing records, page sequence, detention time, and browsing frequency. Different techniques have been employed to segment customers, such as clustering and analyzing shop visits based on factors like time spent on each page and visited brands (Moe 2003), separating customers into clusters based on their purchasing frequency and money spent (Chen and Cheng 2009), and utilizing e-shop characteristics and shopping motivation to identify distinctive shopper subgroups (Ganesh et al. 2010). Other approaches include fuzzy co-clustering (Rathipriya and Thangavel 2010) to recognize behavior patterns on specific web pages, calculating customer interest rates based on browsing time (Zheng et al. 2010), and designing soft-clustering algorithms (R. S. Wu and Chou 2011) to segment customers using multi-category data such as satisfaction, behavior, internet usage level, and demographics. One study developed a rough leader-based clustering procedure (Su and Chen 2015) using customer behavioral data to discover and segment their interest patterns. While understanding customers' behavior is important for sellers to establish connections and improve their e-commerce platform, the focus of the aforementioned studies was solely on segmentation techniques. The following paragraphs will review related works that provide recommendations for e-shop customers based on their behavior data.

In (Kim et al. 2005), researchers analyzed the behavioral and navigational patterns of consumers to estimate their interest levels in products. They developed a collaborative filtering (CF) method that made recommendations based on the estimated preference levels derived from these patterns. Building upon the work of Kim et al. (2005), (Kim and Yum 2011) further improved the estimation of preference levels by integrating three confidence states based on visited products, products added to the cart, and purchased products. They also employed association rule mining to uncover the inter-associations of multiple products included in a customer's interactions. Another recommendation technique was proposed in (K. C. Lee and Kwon 2008), which utilized causal maps and the elaboration likelihood model to enhance performance by incorporating qualitative factors into the recommendation process. In (C. L. Huang and Huang 2009), sequential patterns with time decomposition were leveraged to design a novel CF method that accurately understood and forecasted customers' purchase patterns. This method introduced a double-phase recommendation approach for both the product category and specific products, enabling the prediction of consumer purchasing behavior.

In (Choi et al. 2012), the authors applied clients' purchase patterns identified through sequential pattern analysis (SPA) in conjunction with the regular CF-based method to develop a hybrid recommendation approach. They demonstrated that merging the CF-based method and SPA-based method improved recommendation performance compared to traditional CF approaches. Authors in (Abdullah et al. 2013) proposed a user profiling procedure based on customers' click streams in e-commerce to determine their preferences and accurately identify their neighbors. They combined the CF method with a search-based method, presenting a procedure to generate an aggregated query based on the initial query of the active customer and the preferences of their neighbors, thereby enhancing the quality of recommendations. In (Adeniyi et al. 2014), a real-time recommendation approach was devised to provide personalized recommendations to customers based on their behavioral patterns inferred from click information. The aim was to offer fitting recommendations without the need for explicit requests. The K-Nearest-Neighbor technique was employed to classify customers' click-stream data, assigning them to specific neighbor groups and suggesting customized items that fulfill individual customer requirements at different times. In (Liao and Chang 2016), the authors proposed an association rule method using rough set analysis to analyze customers' preferences. The proposed approach leveraged the analytic hierarchy process for processing ordinal data scales. It effectively generated association rules and adapted them to provide recommendations.

In (Rawat et al. 2017), customer behaviors captured from click-stream data were utilized to predict product preferences, both for items that have already been consumed and those not yet experienced by customers. The CF technique was applied to predict preferences for unexplored products, while the proximity significance singularity, in conjunction with the rough set clustering method, was incorporated in the CF process to enhance the recommendation performance. Furthermore, in (Xiao and Ezeife 2018), the authors combined shopping frequency data with the consequential associations between sequences of clicks and purchases. This integration aimed to enrich the user-item rating matrix, considering both quality and quantity aspects, thus improving the recommendation system. In (Iwanaga et al. 2019), the authors addressed the challenge of calculating a high-quality rating matrix by leveraging clickstream data to improve CF-based recommendations. They introduced a shape-restricted optimization model to estimate item ratings based on the frequency and recency of customers' previous interactions with the product. The authors demonstrated that utilizing the proposed high-quality rating matrix enhances the effectiveness of various CF-based recommendation techniques that rely on the rating matrix. The study (Nishimura et al. 2020) focused on analyzing the association between product view records and the decision-making behavior of customers. The authors proposed a shape-restricted optimization approach to precisely determine the probabilities of product choices for each potential sequence of product views. By employing this approach, they aimed to improve the accuracy of predicting customer preferences and choices based on their browsing history.

In (Lei et al. 2022), the authors introduced a framework for sequential recommender systems designed to capture users' evolving interests from their historical behavior and predict their future preferences. Their framework specifically addresses the challenge of modeling sequential user behavior and preferences. By incorporating techniques such as non-invasive embedding, time-aware modeling, and attention-based fusion, the proposed approach enhances the accuracy and effectiveness of the recommendation system. This framework aims to provide improved recommendations by considering the dynamic nature of user preferences over time. In (Duan et al. 2022), the authors presented a novel model known as MhSa-GRU, specifically designed for product recommendation systems. This model combines Multi-head Self-attention with a Gated Recurrent Unit (GRU) and integrates diverse features like prices, user behavior, and item categories to generate recommendations for the next item. Notably, the MhSa-GRU model incorporates an enhanced self-attention layer that effectively captures correlations among items, while multiple attention heads gather comprehensive local information about the vector representation. In (J. Wu et al. 2022), the authors investigated the research gap related to repeated shopping behavior and aimed to enhance the recommendation process. They developed a novel recommendation method that employs a "split and conquer" strategy to suggest products to customers. The authors categorized customers' repetition shopping behaviors into four types: active customers with stable desires, inactive customers with stable desires, active customers with unstable desires, and inactive customers with unstable desires. Their proposed method consists of two main components: first, identifying the different types of repetition shopping behaviors exhibited by customers, and second, generating a product recommendation list tailored to each customer based on their specific repetition shopping behavior. In (Gan et al. 2023), the authors propose a novel framework called FPD. This framework generates embeddings and builds training losses, capturing user preference differences among behaviors. Instead of relying solely on initial embeddings for predictions, an additional supplementary score is computed. Multiple training losses are constructed using different behaviors to optimize model parameters effectively. The authors also introduce personalized positive weights for each user to balance the difference in interaction records.

## 3. Proposed Methods

This study introduces a novel approach, namely Behavior-Based Recommendation using Category-Based Clustering (BRC), specifically tailored for e-shops. The main objective is to develop an innovative clustering method and recommendation methodology based on customers' behavioral patterns. The clustering method, which is elaborated in Subsection 3.1, and the recommendation methodology, discussed in Subsection 3.2, are meticulously described in this research. To provide a visual representation of the proposed BRC system, Figure 3 presents a block diagram, offering an overview of the framework.

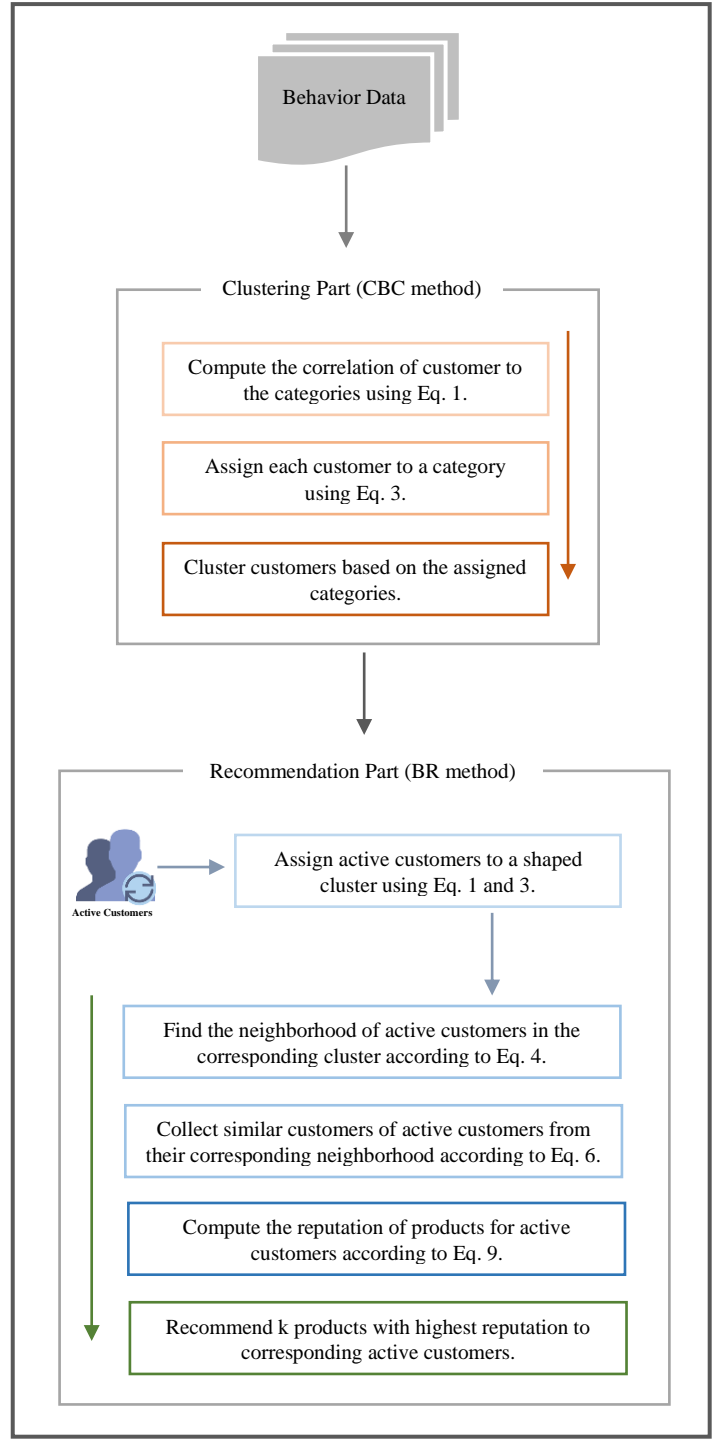

**Figure 3.** The block diagram of BRC method.

### 3.1. Clustering Method

In this research study, a method for clustering users based on their behaviors is presented. The proposed clustering method, known as Category-Based Clustering (CBC), utilizes the system's predefined categories. For instance, consider an e-shop encompassing categories such as smartphones, laptops, headphones, gaming consoles, and chargers. In this context, the proposed CBC technique partitions customers into five distinct groups based on their relevance to these categories. The fundamental aim of this methodology is to identify target customers who have a specific interest in category objects, while also identifying other customers who exhibit similar preferences for category objects.

A fundamental challenge in modeling the CBC lies in determining the appropriate object category for each customer based on their behavior records. To address this challenge, a straightforward mathematical method is proposed to calculate the correlation of customers with respect to the different categories. The correlation of a customer to a particular category is defined as the ratio of their cooperative interactions within that category to their total count of behaviors. To formalize the calculation of correlation, the following equation has been devised:

$$Correlation_{Cust_i}^{Cat_c} = \frac{Cooperation_{Cust_i}^{Cat_c}}{CB_{Cust_i}} \qquad 1$$

$$0 \leq Correlation_{Cust_i}^{Cat_c} \leq 1$$

In the equation provided above, the symbol $Cat_c$ represents a specific category $c$ the set of categories. The symbol $Cust_i$ represents a particular customer $i$ from the set of customers. $CB_{Cust_i}$ denotes the total count of behaviors exhibited by customer $i$. $Cooperation_{Cust_i}^{Cat_c}$ represents the cooperation rate of customer $i$ in category $c$. The cooperation of a customer in a specific category is defined as the summation count of their behaviors within that category. Essentially, cooperation reflects the number of actions performed by customers within a particular category, accounting for various behavior types. The computation for the cooperation level can be expressed using the subsequent equation.

$$Cooperation_{Cust_i}^{Cat_c} = \sum_{p=1}^{P} count\ (Behavior_i, Product_p) \qquad 2$$

$$where \quad Product_p \in Cat_c$$

In equation 2, the variables $Behavior_i$ and $Product_p$ respectively represent the behavior of customer $i$ and product $p$ from product set. The $count\ (Behavior_i, Product_p)$ counts the occurrences of behavior exhibited by customer $i$ on product $p$. It measures how many times customer $i$ has engaged in the specific behavior (e.g., viewed, added to cart, purchased, added to favorites) for product $p$. Let us now delve into a detailed explanation of the CBC method. In the initial step of the CBC process, based on the definition provided above, a customer-category matrix is constructed to represent the correlation between each customer and each category. This matrix is referred to as the correlation matrix. Subsequently, each customer is assigned to the category that exhibits the highest correlation. The mathematical representation of this category assignment can be observed in the following equation.

$$CL_{Cust_i} \xleftarrow{Assign} Cat_c : argmax\ (Correlation_{Cust_i}^{Cat_c});\quad c = 1, 2, \dots, C \qquad 3$$

In the preceding equation, the notation $CL_{Cust_i}$ represents the category label assigned to customer $i$. Once a category has been assigned to each customer, clusters are formed by grouping customers who have been assigned to the same category. It is important to note that the number of clusters is equal to the number of categories. In other words, each category corresponds to a distinct cluster. Additionally, the pseudocode for the CBC method is presented below:

| The pseudocode of CBC method |
|---|
| **Input** |
| $U$: $dataset$ |
| **Output** |
| $CL$ |
| $Clusters$ |
| **Being:** |
| **Step 1: Assigning a category label to each customer** |
| For each $Cust_i$ in $U$ do: |
|   $max = 0$ |
|   For each $Cat_c$ in $U$ do: |
|     Calculate the $Cooperation_{Cust_i}^{Cat_c}$ (using Eq. 2) |
|     Calculate the $Correlation_{Cust_i}^{Cat_c}$ (using Eq. 1) |
|     If $Correlation_{Cust_i}^{Cat_c} > max$: |
|       Assign $Cat_c$ as the $CL_{Cust_i}$ (regarding to Eq. 3) |
| Return $CL$ |
| **Step 2: Shaping clusters** |
| $Clusters_{set} = Categories_{set}$ |
| For each $Cluster_i$ in $Clusters_{set}$ do: |
|   For each $Cust_i$ do: |
|     If $Cluster_i == CL_{Cust_i}$: |
|       $Cluster_i \leftarrow Cust_i$ |
| Return $Clusters$ |

The provided pseudocode elucidates the procedural steps of the CBC method. It initiates by assigning a category label to each customer using the highest correlation criterion and subsequently establishes clusters by associating customers with their corresponding categories.

Overall, the CBC method presents a rigorous and systematic approach for clustering users based on their behaviors. By leveraging preexisting categories, it enables the e-shop platform to deliver targeted recommendations, enhancing the overall user experience and satisfaction.

### 3.2. Recommendation Methodology

Upon completion of the customer interest modeling using the proposed CBC method, it is necessary to outline the recommended methodology for generating personalized recommendations. The suggested approach, known as Behavior-Based Recommendation (BR), is primarily grounded in customer behaviors. The recommendation process commences upon the arrival of an active customer.

The initial step involves determining the cluster to which the active customer belongs. To accomplish this, the cluster label of the active customer is assigned using a procedure similar to that employed in the first step of the CBC method. Subsequently, the active customer's correlation with different categories is calculated, and the category displaying the highest correlation is designated as the cluster label for the active customer. Consequently, the active customer is associated with a specific cluster. Consequently, this enables the recommendation of products within the same category to the active customer, as well as to those customers who have similar preferences.

Next, a neighborhood is formed for the active customer within the identified cluster. To achieve this, a straightforward approach termed Neighborhood Formation is employed in this study. In this process, co-cluster customers who have exhibited sufficient behaviors within the cluster, similar to those of the active customer, comprise the neighborhood for that particular customer. The formulation of Equation 4, provided below, facilitates the identification of the active customer's neighborhood.

$$Neighborhood_{Cust_i} \xleftarrow{Join} Cust_j : CB_{Cust_j} \geq \delta_{Cust_i}$$
$$where \ \ CL_{Cust_i} == CL_{Cust_j} \ and \ \ j = 1, 2, \ldots, J \qquad 4$$

$$\delta_{Cust_i} = round\left(\frac{1}{2}(Max\_CB^{CL_i} + CB^{CL_i}_{Cust_i})\right) \qquad 5$$

Within the aforementioned equations, $\delta_{Cust_i}$ represents the threshold value that determines whether a co-cluster customer can join the neighborhood of customer $i$ (It should be noted that the value of $\delta_{Cust_i}$ is rounded down). $Max\_CB^{CL_i}$ denotes the maximum count of behaviors observed within cluster $i$. $CB^{CL_i}_{Cust_i}$ represents the count of behaviors exhibited by customer $i$ within the corresponding cluster. The neighborhood is composed of customers whose behavior length falls within the range of frequencies observed in the active customer, up to the maximum behavior length recorded within the cluster. Following the formation of the neighborhood, the similar customers to the active one are identified using the equations provided below.

$$Similar^{Set}_{Cust_i} \xleftarrow{Join} Cust_j : Similarity_{Cust_{i,j}} \geq \vartheta_{Cust_i} \qquad 6$$

$$Similarity_{Cust_{i,j}} = \frac{\bigcup_{p=1}^{P}\left(Behavior^p_{Cust_i} \cap Behavior^p_{Cust_j}\right)}{\bigcup_{p=1}^{P}\left(Behavior^p_{Cust_i} \cup Behavior^p_{Cust_j}\right)}, \ \ Cust_j \in Neighborhood_{Cust_i} \qquad 7$$

$$\vartheta_{Cust_i} = mean \sum_{j=1}^{n} Similarity_{Cust_{i,j}} \qquad 8$$

Within the equations mentioned above, $Behavior^p_{Cust_i}$ represents the behaviors exhibited by customer $i$ on product $p$. It is worth noting that when measuring similarity (Eq. 7), the intersection between customers is calculated for each behavior type at each product. This includes behaviors such as viewing, adding to cart, purchasing, and adding to favorites. $Similarity_{Cust_{i,j}}$ denotes the similarity value between customers $i$ and $j$, calculated using the Jaccard equation based on their behaviors. $\vartheta_{Cust_i}$ represents the similarity threshold value for customer $i$. Customers within the neighborhood of the active customer are considered similar if their similarity value is greater than or equal to $\vartheta$. $Similar^{Set}_{Cust_i}$ indicates the set of similar customers for customer $i$.

Once the similar customers of the active customer are identified, a reputation rank is established for products based on the behaviors exhibited by customers within the similar set, considering the influence of similarity. In this study, reputation is defined by the following equation (Equation 9).

$$Reputation^p_{Cust_i} = \sum_{j=1}^{J} count(Behavior_{Cust_j}, Product_p) \times Similarity_{Cust_{i,j}} \qquad 9$$
$$where \ \ Cust_j \in Similar^{Set}_{Cust_i}$$

Equation 9 sums up the product of the count of behavior occurrences for each similar customer and the corresponding similarity value. This calculation allows us to determine the reputation of product $p$ for customer $i$ based on the behaviors of similar customers. Subsequently, products with a higher reputation value are recommended to the active users. Moreover, the pseudocode for the Behavior-Based Recommendation (BR) process is provided below:

**The pseudocode of the BR method**

**Input**

$U$: dataset
$CL$
$Clusters$
$AC$: Active Customers

**Output**

$Recommendations$

**Being:**

**Step 1: Assigning the active customers to a shaped cluster**

$max = 0$

For each $Cust_i$ in $AC$ do:
    For each $cl_c$ in $CL$ do:  (**Note.** $cl_c == Cat_c$ regarding to that each category is indication of a cluster label)
        Calculate the $Cooperation_{Cust_i}^{cl_c}$ (using Eq. 2)
        Calculate the $Correlation_{Cust_i}^{cl_c}$ (using Eq. 1)
        If $Correlation_{Cust_i}^{cl_c} > max$:
            Assign $cl_c$ as the $CL_{Cust_i}$ (regarding to Eq. 3)
    Join the $Cust_i$ to the $Cluster_{cl_c}$

**Step 2: Forming the neighborhood**

Set the $MLB^{cl_c}$
Set the $LB_{Cust_i}$
Set $\alpha$ and $\beta$
Calculate $\delta_{Cust_i}$

For each $Cust_j$ in $Cluster_{cl_c}$ do:
    If $LB_{Cust_j} \geq \delta_{Cust_i}$:
        $Neighborhood_{Cust_i} \xleftarrow{Join} Cust_j$

**Step 3: Finding the similar set**

For each $Cust_j$ in $Neighborhood_{Cust_i}$ do:
    Calculate $Similarity_{Cust_{i,j}}$ (using Eq. 7)

Calculate $\vartheta_{Cust_i}$ (using Eq. 8)

For each $Cust_j$ in $Neighborhood_{Cust_i}$ do:
    If $Similarity_{Cust_{i,j}} \geq \vartheta_{Cust_i}$:
        $Similar_{Cust_i}^{Set} \xleftarrow{Join} Cust_j$

**Step 4: Recommendation**

$Recommendations = [\ ]$
$Temporary\ suggestion\ list = [\ ]$

For each $Product_p$ in $U$ do:
    $Reputation_{Cust_i}^p = 0$
    For each $Cust_j$ in $Similar_{Cust_i}^{Set}$ do:
        $Reputation_{Cust_i}^p \mathrel{+}= Count(Behavior_{Cust_j}, Product_p) \times Similarity_{Cust_{i,j}}$
    $Temporary\ suggestion\ list \leftarrow Reputation_{Cust_i}^p$

    $Recomendations_{Cust_i} \leftarrow k\ product$ with highest $Reputation$ from the $Temporary\ suggestion\ list$

Return $Recommendations$

The provided pseudocode presents the step-by-step process of the BR approach, including assigning the active customers to a cluster, forming the neighborhood, finding the similar set, and making recommendations based on reputation ranks.

Overall, this methodology aims to generate personalized recommendations by considering the behavior patterns of customers and leveraging similarities with other customers in the same cluster. By utilizing the reputation of products based on similar customers' behaviors, the approach provides recommendations that align with the active customer's preferences.

## 4. Experimental Results and Analysis

This section provides a comprehensive presentation of the experiments conducted in this study, along with the corresponding results, to demonstrate the performance of the proposed method in comparison to benchmark approaches. The section is structured as follows: Subsection 4.1 introduces the dataset utilized in this case study. Subsection 4.2 describes the evaluation measures applied in this study. Subsection 4.3 introduces the state-of-the-art methods utilized for comparison purposes. Subsection 4.4 explains the simulation methodology employed in this study. Subsection 4.5 presents and analyzes the experimental outcomes.

### 4.1. Dataset

For the experiments in this study, a dataset of user behaviors from the Taobao online shopping platform, provided by Alibaba Group, was utilized. The dataset can be accessed via the link "User Behaviors Dataset". It encompasses over one hundred million user activities, involving 987,994 unique users, across approximately four million products and 9,439 distinct categories. The dataset provides the following information: user ID, category ID, item ID, behavior types, and timestamps. The behavior types are categorized into four modes, as outlined in Table 1. Additionally, Table 2 presents the statistical information of the dataset.

**Table 1.** The behavior types existed in the dataset.

| Behavior Types | Explanations |
| --- | --- |
| Pv | Viewing a product by a user |
| Fav | Adding a product into the favorites list by a user |
| Cart | Adding a product into the shopping cart by a user |
| Buy | Purchasing a product by a user |

**Table 2.** The statistical information of the dataset.

| Title | Statistical information |
| --- | --- |
| Customer Id | 987,994 |
| Category Id | 9,439 |
| Product Id | 4,162,024 |
| Recorded Behaviors | 100,150,807 |

### 4.2. Evaluation Measures

For evaluating the clustering performance, two widely recognized criteria, namely the Davies-Bouldin index (DB) and the Dunn index, were utilized. These indices can be computed using Equations 10 and 11.

$$DB = \frac{1}{N} \sum_{i=1}^{N} \max_{i \neq j} \frac{\Delta(c_i) + \Delta(c_j)}{S(c_i, c_j)} \tag{10}$$

$$Dunn = \frac{\max_{i,j=1,\dots,C; i \neq j} S(c_i, c_j)}{\min_{i=1,\dots,C} \Delta(c_i)} \tag{11}$$

In the above equations, $S(c_i, c_j)$ represents the similarity between clusters $c_i$ and $c_j$ (inter-cluster similarity). $\Delta(c_i)$ and $\Delta(c_j)$ denote the intra-cluster similarity of clusters $c_i$ and $c_j$, respectively. $N$ denotes the number of clusters. The objective of both the DB index and Dunn index is to identify cluster sets that are dense and well-separated. A high DB value and a small Dunn value indicate that the derived clusters are compact, with their centers being distant from each other (Su and Chen 2015).

To evaluate the recommendation performance, three well-established metrics, namely precision, recall, and F-measure, were employed. These metrics can be calculated using Equations 12, 13, and 14.

$$Precision = \frac{1}{A} \sum_{i=1}^{A} \frac{1}{K} (|Rec_{Cust_i}^{K} \cap P_{Cust_i}^{Rel}|) \tag{12}$$

$$Recall = \frac{1}{A} \sum_{i=1}^{A} \frac{1}{P_{Cust_i}^{Rel}} (|Rec_{Cust_i}^{K} \cap P_{Cust_i}^{Rel}|) \tag{13}$$

$$F_{Measure} = \frac{2 * Precision * Recall}{Precision + Recall} \tag{14}$$

In the above equations, $A$ represents the number of active customers who received recommendations, while $K$ denotes the count of provided recommendations. $Rec_{Cust_i}^{K}$ and $P_{Cust_i}^{Rel}$ respectively represent the $K$ recommended products for customer $i$ and the relevant products of customer $i$. Precision and recall serve as metrics for exactness and completeness, respectively. The F-measure balances the trade-off between the precision and recall rankings (Kim and Yum 2011; Liao and Chang 2016).

### 4.3. Comparison Methods

Since the study evaluates the proposed methods in two segments, clustering and recommendation, two types of methods are compared: clustering methods and recommendation methods.

To assess the effectiveness of the proposed CBC method, two baseline clustering methods, fuzzy c-means (FCM) and k-means, are employed for comparison. Additionally, a state-of-the-art method called Rough-leader clustering method (RL), proposed in (Su and Chen 2015) for clustering behavioral data similar to the dataset used in this study, is employed to demonstrate the significant performance of the CBC method. Furthermore, modified versions of the proposed BRC method are designed by applying FCM, k-means, and RL clustering methods instead of the CBC method. These modified methods are referred to as FCM-BR method, Kmeans-BR method, and RL-BR method. By comparing the proposed BRC with these modified methods, the efficiency of the proposed CBC method in recommendations against FCM, k-means, and RL clustering methods can be demonstrated.

For comparing the performance of the proposed BRC in recommendation, the following state-of-the-art methods are applied:

- **CSD-RS** (Kim and Yum 2011): An advanced baseline method that focuses on the confidence metric. CSD-RS estimates confidence levels between viewed products, products added to the cart, and bought products, and then determines the preference ranking through a linear integration of these levels. Finally, it recommends the top-N products to the customer.

- **ARS-GB** and **ARS-RF** (Rawat et al. 2017): These methods utilize clickstream data to predict customer preference values for products that were clicked but not purchased, using efficient classifiers such as random forest (ARS-RF) and gradient boosting (ARS-GB). Collaborative filtering (CF) procedures are then implemented to provide product recommendations.

- **HPCRec** (Xiao and Ezeife 2018): This method employs customers' clickstream data to enrich the rating matrix by considering both quantity (modifying primary ratings of zero to consequential bond values) and quality (modifying primary ratings of one to a value between zero and one that incorporates the shopping frequency of a product). The enriched matrix is then utilized in the CF procedure to generate recommendations.

By comparing the proposed BRC method with these state-of-the-art methods, its performance in recommendation can be evaluated effectively.

### 4.4. Simulation Explanation

The simulation conducted in this study aimed to address the challenges posed by the large and sparse dataset, as indicated in Table 2. To facilitate the simulation process and ensure its effectiveness, a two-phase preprocessing approach was employed. The details of the preprocessing phases are outlined below.

**First Phase:** In this phase, products that exhibited significant contribution across the four behavioral modes were selected as the product collection. Consequently, products that solely had "Pv" (product viewing) operations and had a number of "Pv" occurrences below a threshold value, denoted as $\varphi = 1000$, were filtered or excluded from the dataset. Multiple iterations were performed to identify the best conditions that minimized the loss of valuable information. The statistical information of the dataset after the first preprocessing phase is provided in Table 3 and can be compared with the initial statistical information.

Table 3. The statistical information after the first phase of preprocessing compared to the basic statistical information of dataset.

| Title | Statistical information in the dataset | Statistical information after the first phase of filtering |
|---|---|---|
| Customer Id | 987,994 | 966,045 |
| Category Id | 9,439 | 8,207 |
| Product Id | 4,162,024 | 884,226 |
| Recorded Behaviors | 100,150,807 | 20,863,640 |

Based on the information in Table 3, it is observed that 3,277,798 products that had only been viewed ("Pv") were excluded from the dataset after the first phase. Additionally, there were 1,232 categories where all their subset products had only been viewed, leading to their exclusion from the dataset. Furthermore, 21,949 customers who solely viewed the filtered products and did not engage in any other actions were also excluded. As a result, the first preprocessing phase reduced the number of behavior records from 100,150,807 to 20,863,640.

**Second Phase:** In contrast to the first phase, which focused on filtering products, the second phase of preprocessing centered around customers. Customers who demonstrated significant contribution based on the count of their behavior records were selected as part of the customer collection. A threshold value, denoted as $\tau$, was defined to exclude customers with behavior records below this threshold. In this case, $\tau$ was set to 50, considering that the maximum count of recorded behaviors for a single customer in the dataset was 250. Only one condition was defined to exclude customers: the number of behavior records must be less than the threshold $\tau=50$. The statistical information after performing the second preprocessing phase is presented in Table 4 and can be compared to the achieved statistical information from the first preprocessing phase.

Table 4. The statistical information after the second phase of preprocessing compared to statistical information of the first phase.

| Title | Statistical information after the first phase of filtering | Statistical information after the second phase of filtering |
|---|---|---|
| Customer Id | 966,045 | 5,748 |
| Category Id | 8,207 | 4,686 |
| Product Id | 884,226 | 162,127 |
| Recorded Behaviors | 20,863,640 | 665,063 |

Based on the information in Table 4, it is observed that 960,297 customers with behavior records below the threshold $\tau$ were excluded from the dataset during the second preprocessing phase. Additionally, this phase resulted in reducing the number of products to 162,127 from 884,226, and the number of categories to 4,686 from 8,207. Consequently, the behavior records were reduced to 665,063 from 20,863,640. This preprocessing step facilitated the application of the dataset for conducting the simulation.

Considering that the experiments were conducted in two phases, namely clustering and recommendation performance, two simulations were carried out for evaluation, as explained below:

- Clustering Evaluation Simulation: The CBC method, along with the comparison methods of clustering, was evaluated on the preprocessed data described above. The CBC method and the aforementioned clustering comparison methods were applied to cluster customers. The resulting clusters were evaluated using the DB index and Dunn index measures. The results are presented and compared in Subsection 4.5.1.

- Recommendation Evaluation Simulation: In this simulation, the proposed BRC method was applied to the entire preprocessed dataset to evaluate its performance comprehensively. Initially, the data was divided into a testing set (20%) and a training set (80%) using the K-fold cross-validation method, where K=10 in this case. In the clustering part of BRC, the CBC method was applied to the training data to cluster customers. In the recommendation part of BRC, the BR method was applied to the testing data to identify the cluster to which each test customer belongs. Subsequently, neighborhoods were formed for each test customer, and similar customers were collected from these neighborhoods. The reputation of products for each test customer was estimated based on the collected similar customers, and products with higher reputation values were recommended. The compared methods were also applied following a similar procedure. The results of this evaluation are presented and compared in Subsection 4.5.2.

### 4.5. Experiments Outcomes

### 4.5.1. Clustering Evaluation Results

During the initial stage, the evaluation outcomes of the proposed CBC method are meticulously presented and contrasted against three algorithms, namely RL, FCM, and K-means. The summarized results can be found in Table 5. The proposed CBC method showcases remarkable clustering performance, as evidenced by DB and Dunn indexes of 46.6 and 1.7473, respectively. Among the compared methods, RL achieves better results for the DB and Dunn indexes, with values of 34.21118 and 3.4137, respectively, following the CBC method. Consequently, the CBC method enhances clustering performance by 12.38 and 1.66 based on the DB and Dunn indexes, respectively, in comparison to RL. As for FCM and K-means, the results are provided with varying cluster numbers due to the necessity of setting the partition number during initialization. FCM attains the optimal outcomes for the DB and Dunn indexes, obtaining values of 27.2298 and 5.485844, correspondingly, while utilizing 82 clusters. Thus, the CBC method improves clustering performance by 19.37 and 3.73 for the DB and Dunn indexes, respectively, in comparison to FCM. Similarly, K-means achieves the best results for the DB and Dunn indexes, specifically 26.8541 and 5.9622, respectively, with 82 clusters. Consequently, the CBC method enhances clustering by 19.74 and 4.21 for the DB and Dunn indexes, respectively, relative to K-means. Overall, the proposed CBC method significantly enhances clustering in relation to its competitors in this simulation.

**Table 5.** Comparison results of DB and Dunn indexes for the proposed CBC and baseline methods.

| Method | K | DB index | Dunn index |
| --- | --- | --- | --- |
| CBC (proposed) | 4,686 | **46.6** | **1.7473** |
| RL | 153 | 34.21118 | 3.4137 |
| FCM | 45 | 15.741 | 11.5244 |
| K-means | 45 | 14.638 | 12.37441 |
| FCM | 66 | 18.44 | 9.79211 |
| K-means | 66 | 17.3921 | 10.84132 |
| FCM | 82 | 27.2298 | 5.485844 |
| K-means | 82 | 26.8541 | 5.9622 |
| FCM | 100 | 23.942 | 7.6821 |
| K-means | 100 | 22.625 | 8.2473233 |

In the subsequent stage, the efficacy of the CBC method in enhancing recommendation quality is assessed in comparison to FCM, k-means, and RL. To accomplish this, three modified methods of the proposed recommendation methodology, namely FCM-BR, Kmeans-BR, and RL-BR, are devised, as discussed in subsection 4.3. Table 6 provides the evaluation results for top-3, top-10, and top-17 recommendations. The findings clearly indicate that the BRC method, which employs the proposed CBC clustering method, exhibits a substantial improvement in performance compared to the modified methods of the proposed recommendation methodology. Specifically, in terms of precision, the BRC method achieves the highest result for top-3 recommendations, attaining a value of 82.6. In contrast, the FCM-BR, Kmeans-BR, and RL-BR methods achieve their best results with values of 70.66, 70.4, and 73.62, respectively. Regarding the F-measure, the BRC method demonstrates the highest efficacy for top-17 recommendations, with a value of 56.18, whereas the RL-BR, FCM-BR, and Kmeans-BR methods achieve values of 43.45, 41.47, and 41.91, respectively.

**Table 6.** The results for comparing proposed CBC method with FCM, k-means, and RL methods in enhancing recommendation performance.

| No. Reco | Method | Precision | Recall | F-measure |
|---|---|---|---|---|
| $Top-3$ | FCM-BR | 70.66 | 26.84 | 38.9 |
| | Kmeans-BR | 70.4 | 26.73 | 38.74 |
| | RL-BR | 73.62 | 26.9 | 39.4 |
| | BRC (proposed) | **82.6** | **35.08** | **49.24** |
| $Top-10$ | FCM-BR | 65.29 | 29.46 | 40.6 |
| | Kmeans-BR | 64.71 | 28.82 | 39.87 |
| | RL-BR | 68.12 | 30.05 | 41.7 |
| | BRC (proposed) | **79.24** | **40.9** | **53.95** |
| $Top-17$ | FCM-BR | 60.5 | 32.07 | 41.91 |
| | Kmeans-BR | 59.93 | 31.71 | 41.47 |
| | RL-BR | 62.87 | 33.2 | 43.45 |
| | BRC (proposed) | **72.81** | **45.74** | **56.18** |

Figure 4 illustrates the average results for these three recommendation sizes. The BRC method achieves notable averages of 78.21 and 40.57 for precision and recall, respectively, surpassing the performance of the other modified methods significantly. In terms of the F-measure, which captures the balance between precision and recall, the BRC method achieves an average of 53.12, while the FCM-BR, Kmeans-BR, and RL-BR methods achieve values of 40.47, 40.02, and 41.51, respectively. A comparison of the average results reveals that the CBC method improves recommendation performance by approximately 13% more than the FCM and K-means methods, and 11% more than the RL method. The results presented in Table 6 and Figure 4 provide clear evidence of the superiority of the CBC method, as employed in the BRC method, in enhancing recommendation performance compared to the other aforementioned clustering methods.

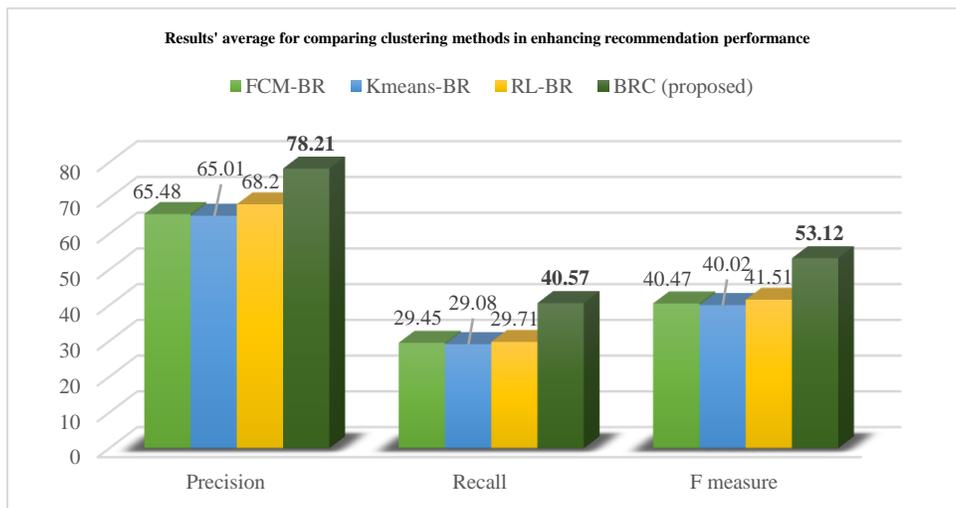

**Figure 4.** The results for comparing proposed CBC method with FCM, k-means, and RL methods in enhancing recommendation performance.

### 4.5.2. Evaluation Results for Recommendations

To assess the performance of recommendations, this study considers two recommendation sizes: small and medium. As part of the evaluation for small-sized recommendations, Table 7 provides a detailed comparison of the results for top-1, top-3, and top-6 recommendations. The proposed BRC method exhibits exceptional performance surpassing all other methods. It achieves the highest precision of 84.53 for top-1 recommendations, along with a recall of 32.61 and an F-measure of 47.06. These outcomes underscore the BRC method's ability to offer precise recommendations, highlighting its superiority in recommendation accuracy. Similarly, for top-3 recommendations, the BRC method continues to outperform alternative approaches, attaining a precision of 82.6, recall of 35.08, and F-measure of 49.24. Furthermore, for top-6 recommendations, the BRC method remains superior, achieving a precision of 80.19, recall of 39.32, and F-measure of 52.76. An examination of the results presented in Table 7 confirms the unequivocal superiority of the BRC method over state-of-the-art techniques such as HPCRec, ARS-GB, ARS-RF, and CSD-RS, specifically for the small size of recommendations. The BRC method consistently outperforms these approaches, establishing itself as a leading solution within the realm of recommendation systems for smaller recommendation sizes.

**Table 7.** The comparison results of small sizes recommendations for proposed BRC and state-of-art methods.

| No. Reco | Method | Precision | Recall | F-measure |
|---|---|---|---|---|
| $Top-1$ | HPCRec | 69.61 | 28.42 | 40.361 |
| | ARS-GB | 66.27 | 26.54 | 37.9 |
| | ARS-RF | 67.49 | 27.22 | 38.793 |
| | CSD-RS | 65.14 | 26.68 | 37.855 |
| | BRC (proposed) | **84.53** | **32.61** | **47.06** |
| $Top-3$ | HPCRec | 67.92 | 28.67 | 40.32 |
| | ARS-GB | 64.137 | 27.1 | 38.1 |
| | ARS-RF | 64.79 | 28.42 | 39.5 |
| | CSD-RS | 63.393 | 27.61 | 38.466 |
| | BRC (proposed) | **82.6** | **35.08** | **49.24** |
| $Top-6$ | HPCRec | 65.233 | 29.148 | 40.292 |
| | ARS-GB | 61.56 | 28.94 | 39.371 |
| | ARS-RF | 62.044 | 29.37 | 39.867 |
| | CSD-RS | 60.827 | 28.79 | 39.08 |
| | BRC (proposed) | **80.19** | **39.32** | **52.76** |

For ease of comparison, Figure 5 provides the average results of each approach for the small size recommendations. Notably, the BRC method achieves the highest average precision, recall, and F-measure of 82.44, 35.67, and 49.68, respectively. Following closely, the state-of-the-art method HPCRec demonstrates average results of 67.58, 28.74, and 40.32 for precision, recall, and F-measure. Thus, compared to HPCRec, the BRC method outperforms with improvements of 14.86%, 6.93%, and 9.36% in precision, recall, and F-measure, respectively. ARS-RF achieves an average precision, recall, and F-measure of 64.77, 28.33, and 39.38, respectively. Consequently, the BRC method surpasses ARS-RF by 17.67%, 7.34%, and 10.3% in terms of recommendation performance. Similarly, ARS-GB exhibits averages of 63.98, 27.52, and 38.45 for precision, recall, and F-measure. Therefore, the BRC method outperforms ARS-GB by up to 18.46%, 8.15%, and 11.23% in precision, recall, and F-measure, respectively.

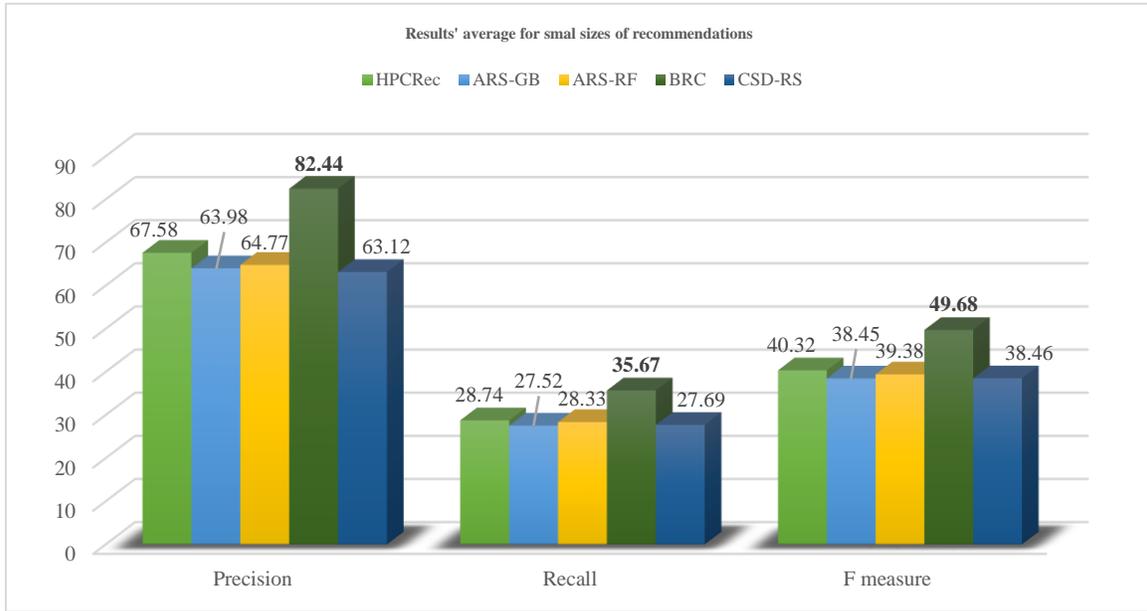

**Figure 5.** The comparison of results' average for small sizes of recommendations.

Additionally, the baseline method CSD-RS achieves averages of 63.12, 27.69, and 38.46 for precision, recall, and F-measure. Consequently, the BRC method significantly improves precision, recall, and F-measure compared to CSD-RS by 19.32%, 7.98%, and 11.22%, respectively. Overall, Figure 5 illustrates the superior efficiency of the BRC method for recommendations in comparison to the aforementioned state-of-the-art methods within the small size recommendation category.

For medium-sized recommendations, which involve suggesting top-10, top-14, and top-17 products to customers, the evaluation results are provided in Table 8. In terms of top-10 recommendations, the proposed BRC method exhibits the highest precision of 79.24, indicating its proficiency in delivering accurate recommendations. Furthermore, BRC surpasses other methods, namely HPCRec, ARS-GB, ARS-RF, and CSD-RS, by achieving superior recall and F-measure scores of 40.9 and 53.95, respectively, which are comparatively lower in the alternative methods. For the top-14 recommendations, the BRC method maintains its superiority with the highest precision, recall, and F-measure scores of 75.52, 43.62, and 55.29, respectively, among all methods. Similarly, in the case of top-17 recommendations, the BRC method excels by demonstrating the highest precision score of 72.81. Additionally, BRC achieves superior recall and F-measure scores of 45.74 and 56.18, respectively, outperforming other methods. In comparison, the alternative methods exhibit lower precision, recall, and F-measure scores. A comprehensive examination of the results presented in Table 8 unambiguously establishes the superiority of the proposed BRC method over state-of-the-art techniques, including HPCRec, ARS-GB, ARS-RF, and CSD-RS, for medium-sized recommendations as well.

Furthermore, Figure 6 presents the average results for medium-sized recommendations. The BRC method achieves the highest average precision, recall, and F-measure scores of 75.85, 43.42, and 55.14, respectively. In contrast, the HPCRec method obtains average scores of 59.55, 31.228, and 40.96 for precision, recall, and F-measure. As a result, the BRC method outperforms HPCRec by 16.3%, 12.19%, and 14.18% in terms of precision, recall, and F-measure, respectively. Similarly, compared to the ARS-RF method, the BRC method shows improvements of 17.37%, 12.2%, and 14.44% in precision, recall, and F-measure. The ARS-GB method achieves average scores of 57.69, 30.49, and 39.88 for precision, recall, and F-measure. In contrast, the BRC method surpasses ARS-GB by 18.16%, 12.93%, and 15.26% in precision, recall, and F-measure, respectively. Additionally, the baseline CSD-RS method attains average scores of 57.21, 30.12, and 39.46 for precision, recall, and F-measure. In comparison, the BRC method demonstrates substantial improvements over CSD-RS, surpassing it by 18.64%, 13.3%, and 15.68% in precision, recall, and F-measure, respectively. Figure 6 provides additional evidence of the BRC method's superior effectiveness in recommending relevant products compared to state-of-the-art methods.

**Table 8.** The comparison results of medium sizes recommendations for proposed BRC and state-of-art methods.

| No. Reco | Method | Precision | Recall | F-measure |
|---|---|---|---|---|
| $Top-10$ | HPCRec | 61.57 | 31.275 | 41.479 |
| | ARS-GB | 59.76 | 30.4 | 40.299 |
| | ARS-RF | 60.8 | 31.35 | 41.369 |
| | CSD-RS | 59.133 | 30.212 | 39.99 |
| | BRC (proposed) | **79.24** | **40.9** | **53.95** |
| $Top-14$ | HPCRec | 59.241 | 31.91 | 41.47 |
| | ARS-GB | 57.389 | 31.347 | 40.546 |
| | ARS-RF | 58.46 | 32.15 | 41.485 |
| | CSD-RS | 57.12 | 30.68 | 39.918 |
| | BRC (proposed) | **75.52** | **43.62** | **55.29** |
| $Top-17$ | HPCRec | 57.846 | 30.5 | 39.94 |
| | ARS-GB | 55.94 | 29.728 | 38.82 |
| | ARS-RF | 56.19 | 30.16 | 39.25 |
| | CSD-RS | 55.378 | 29.487 | 38.48 |
| | BRC (proposed) | **72.81** | **45.74** | **56.18** |

Figure 7 provides a comprehensive overview by presenting the total average results for both small and medium recommendation sizes, facilitating a conclusive comparison between the proposed BRC method and the state-of-the-art techniques. In terms of precision, the BRC method achieves the highest average score of 79.145%, surpassing HPCRec by 15.57% and CSD-RS by 18.98%. Likewise, for recall, BRC attains an average score of 40.04%, outperforming HPCRec by 10.05% and CSD-RS by 11.13%. Moreover, BRC demonstrates an average F-measure score of 52.41%, exceeding HPCRec and CSD-RS by 11.76% and 13.44%, respectively. The comparison depicted in Figure 7 clearly highlights the superiority of the BRC method over the state-of-the-art methods. Overall, the results indicate that the proposed BRC method consistently outperforms the state-of-the-art methods across all recommendation sizes. It achieves higher precision, recall, and F-measure values, demonstrating its effectiveness in providing accurate and relevant recommendations. These findings suggest that the BRC method is effective in providing accurate and relevant recommendations, making it a promising approach in the field of recommendation systems. Furthermore, the upcoming section provides a discussion on the BRC method and the state-of-the-art methods.

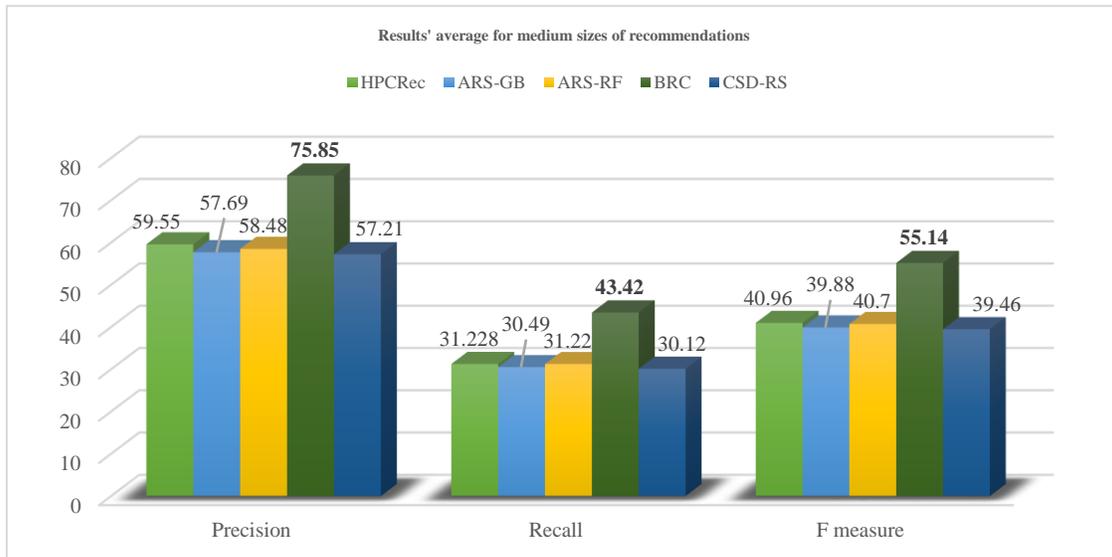

**Figure 6.** The comparison of results' average for medium sizes of recommendations.

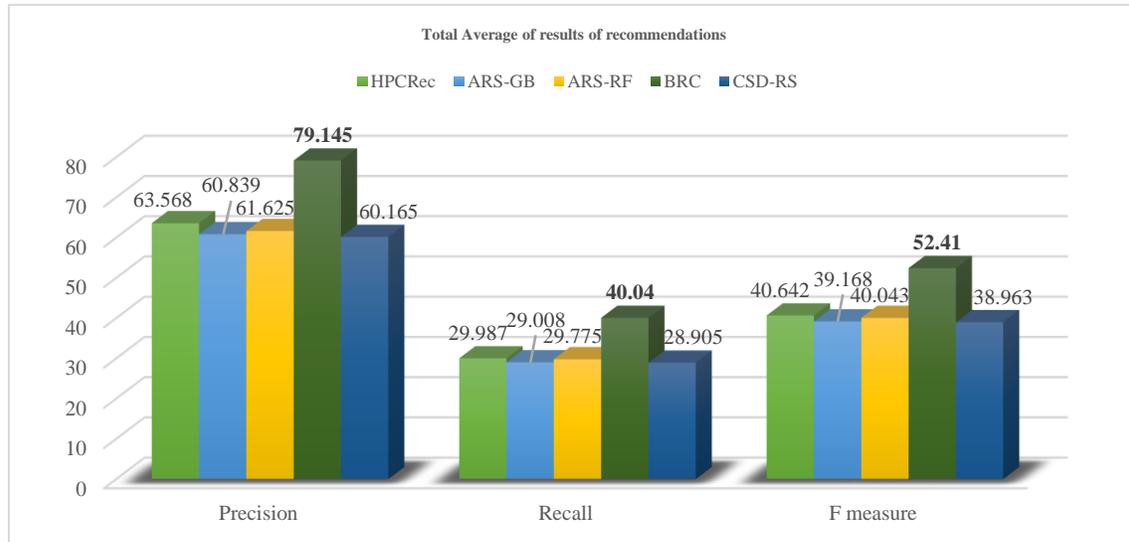

**Figure 7.** Total average of results comparison for recommendations.

## 5. Discussion

This section aims to provide a comprehensive and insightful discussion on the experimental comparison conducted in the study, with a specific focus on two key aspects: the evaluation of results and the contextual analysis. In summary, this section goes beyond a mere presentation of the experimental comparison results. It provides a comprehensive and scholarly discussion that critically evaluates the obtained results, highlights the superiority of the proposed method, and places the study within the broader context of recommendation systems and e-commerce platforms. By combining the evaluation of results with contextual analysis, this section offers valuable insights into the research findings and their implications, contributing to the advancement of knowledge in the field. In accordance with this, the discussion is organized into two subsections, namely the result analysis statement (subsection 5.1) and the contextual analysis statement (subsection 5.2).

### 5.1. Result Analysis Statement

The evaluation of the proposed method in this study involved two technical components: the clustering approach (CBC) and the recommendation methodology (BRC). Two phases were conducted to assess the performance of the proposed method compared to state-of-the-art methods in both clustering and recommendation. In the first phase, the study introduced a category-based clustering approach called CBC, specifically designed for behavior data. CBC was compared with FCM, k-means, and RL clustering methods using the DB index and Dunn index to evaluate its clustering performance. As CBC was also designed to enhance the performance of the proposed recommendation methodology, the state-of-the-art methods were applied in the recommendation process alongside CBC to assess their impact on recommendation improvement. The evaluation results, presented in subsection 4.5.1, demonstrated that CBC achieved significant performance improvements compared to the FCM, k-means, and RL methods in both clustering and recommendation.

In the second phase, the study introduced a recommendation methodology called BRC, based on behavior data and utilizing the proposed CBC method. BRC was compared with four state-of-the-art methods (HPCRec, ARS-GB, ARS-RF, and CSD-RS) using precision, recall, and F-measure metrics to evaluate its performance in recommending relevant products. The recommendation performance was evaluated for different sequences of suggestions, and the results were provided in subsection 4.5.2. The findings indicated that BRC exhibited outstanding efficiency compared to HPCRec, ARS-GB, ARS-RF, and CSD-RS methods in enhancing recommendation. Furthermore, when the proposed recommendation methodology was combined with the aforementioned clustering methods, it consistently outperformed HPCRec, ARS-GB, ARS-RF, and CSD-RS methods in most evaluation cases. For instance, the Kmeans-BR approach of the proposed recommendation methodology demonstrated precision, recall, and F-measure results of 70.4, 26.73, and 38.74 in top-3 recommendations (see Table 6), while HPCRec achieved results of 67.92, 28.67, and 40.32 in the same recommendation size (see Table 7). Thus, the Kmeans-BR method exhibited more favorable outcomes in terms of precision and F-measure compared to HPCRec. Further comparisons can be made using the results presented in Tables 6, 7, and 8 for similar recommendation sizes. The analysis and comparison of these results clearly indicate the superiority of the proposed recommendation methodology over the other methods.

**5.2. Contextual Analysis Statement**

Most recommendation methods rely on numerical data such as ratings, which require users to actively provide ratings for items. However, these methods often encounter challenges related to user non-cooperation in expressing ratings. This issue becomes more pronounced in large e-commerce sites like Alibaba and Amazon, where numerous products from diverse categories and brands are available, making it difficult for users to rate multiple products due to time limitations and the sheer number of options. Therefore, rating-based techniques or similar methods that depend on explicit information provided by users are not effective for generating appropriate recommendations in e-commerce settings.

To address this challenge, utilizing customers' behavioral data is considered one of the best solutions for designing efficient recommendation methods for e-commerce sites. Various approaches based on users' behavioral data have been presented in the literature. However, many of these approaches transform behavior information into rating or frequency matrices and develop rating-based methods. For instance, HPCRec converts customers' purchase behavior into a purchase frequency matrix to measure similarity and provide recommendations. ARS-GB utilizes customers' actions such as viewing and adding to the cart, constructing a preference matrix based on the sequence of product visits until purchase, and then applying the Gradient Boosting method for classification and recommendation. Similarly, ARS-RF uses Random Forests for classification and prediction. These methods and others follow the approach of transforming behavioral data into rating or frequency matrices and apply machine learning techniques for prediction, resulting in computational and time complexity.

In contrast, the proposed BRC method, while being simpler, directly utilizes the type of customer behavior, including viewing, adding to favorites, adding to the cart, and purchasing. By directly using customers' behaviors, BRC increases the reliability of discovering efficient neighbors and providing more satisfying recommendations compared to methods that transform behavior data into frequency or rating matrices, which are implicitly estimated and then used for preference prediction. The results from the criterion-based evaluations (examined in section 4.5) further demonstrate the performance improvement achieved by the proposed BRC method through the direct use of behavioral data.

The benchmark CSD-RS method employs association rule techniques to calculate the probability of a product being purchased based on visits, and it utilizes behavioral data, including visits, cart additions, and purchases, to estimate customers' confidence levels in a product. CSD-RS then leverages this information to estimate preferences and recommend highly reliable products. On the other hand, the proposed BRC approach incorporates a clustering procedure based on product categories, which helps in identifying the category from which a customer seeks a product. By leveraging similar neighbors who are active and interested in the same category, BRC predicts the suitability of unknown products for the target customer and provides appropriate recommendations. In essence, BRC determines the relevant product categories for a customer using the proposed CBC method and generates recommendations within those contexts. This combination significantly contributes to the improvement of the BRC method in providing high-quality recommendations compared to the state-of-the-art methods.

**6. Conclusion and Future Works**

The significance of customer loyalty in the realm of e-commerce has prompted the widespread adoption of RS in various online applications. However, most existing RS designs heavily rely on users' rating matrices for items, which poses a challenge when dealing with products that have limited or no ratings, particularly in large-scale e-shop platforms. To address this issue, this study focuses on developing a robust RS approach for e-shops based on customers' behaviors, which can be readily captured from their web platform interactions.

In this study, a CF-based recommendation methodology and a category-based clustering method have been devised, leveraging the recorded customer behaviors derived from click-stream data. As a result, a novel model-based CF procedure, referred to as BRC (Behavior-based Recommendation with Category-based Clustering), has been developed. BRC has been rigorously evaluated using precision, recall, and F-score measures on a behavior dataset provided by Alibaba Group, and its performance has been compared against relevant state-of-the-art methods.

The experimental outcomes clearly demonstrate the superior efficiency of BRC compared to the state-of-the-art methods. The results also highlight that analyzing customers' behaviors captured from click-stream data can be more effective in developing a robust RS and uncovering customers' interests compared to rating data, which often suffers from sparsity issues. Additionally, the proposed category-based clustering method, known as CBC, has been independently evaluated for its clustering performance. CBC has exhibited superior performance when compared to baseline methods such as FCM, k-means, and the state-of-the-art method known as RL, as evidenced by the obtained results of the DB index and Dunn index.

Moving forward, future research should explore important aspects of customers' behavior, such as the temporal dimension (e.g., time and duration of specific actions or behaviors), the relationship between behavior sequences and purchasing decisions, and the influence of customers on each other. Additionally, addressing the cold-start problem by considering the behaviors captured from click-stream data represents another significant area for future works. By investigating these aspects, researchers can further enhance the understanding and effectiveness of behavior-based recommendation systems, leading to advancements in the field of e-commerce.

## References

Abdullah, N., Xu, Y., & Geva, S. (2013). Integrating collaborative filtering and matching-based search for product recommendations. *Journal of Theoretical and Applied Electronic Commerce Research*, *8*(2), 34–48. https://doi.org/10.4067/S0718-18762013000200004

Adeniyi, D. A., Wei, Z., & Yongquan, Y. (2014). Automated web usage data mining and recommendation system using K-Nearest Neighbor (KNN) classification method. *Applied Computing and Informatics*, *12*(1), 90–108. https://doi.org/10.1016/J.ACI.2014.10.001

Alfian, G., Ijaz, M. F., Syafrudin, M., Syaekhoni, M. A., Fitriyani, N. L., & Rhee, J. (2019). Customer behavior analysis using real-time data processing: A case study of digital signage-based online stores. *Asia Pacific Journal of Marketing and Logistics*, *31*(1), 265–290. https://doi.org/10.1108/APJML-03-2018-0088/FULL/XML

Alrezaamiri, H., Ebrahimnejad, A., & Motameni, H. (2019). Solving the next release problem by means of the fuzzy logic inference system with respect to the competitive market. *https://doi.org/10.1080/0952813X.2019.1704440*, *32*(6), 959–976. https://doi.org/10.1080/0952813X.2019.1704440

Ambulgekar, H. P., Pathak, M. K., & Kokare, M. B. (2019). A Survey on Collaborative Filtering: Tasks, Approaches and Applications. *Advances in Intelligent Systems and Computing*, *811*, 289–300. https://doi.org/10.1007/978-981-13-1544-2_24

Barzegar Nozari, R., Koohi, H., & Mahmodi, E. (2020). A Novel Trust Computation Method Based on User Ratings to Improve the Recommendation. *International Journal of Engineering*, *33*(3), 377–386. https://doi.org/10.5829/IJE.2020.33.03C.02

Barzegar Nozari, Reza, & Koohi, H. (2020). A novel group recommender system based on members' influence and leader impact. *Knowledge-Based Systems*, *205*, 106296. https://doi.org/10.1016/J.KNOSYS.2020.106296

Barzegar Nozari, Reza, & Koohi, H. (2021). Novel implicit-trust-network-based recommendation methodology. *Expert Systems with Applications*, *186*, 115–709. https://doi.org/10.1016/J.ESWA.2021.115709

Barzegar Nozari, Reza, & Koohi, H. (2022). An Implicit Trust-Network construction approach and a Recommendation Methodology for recommender systems. *Software Impacts*, *12*, 100242. https://doi.org/10.1016/J.SIMPA.2022.100242

Bedi, P., & Sharma, R. (2012). Trust based recommender system using ant colony for trust computation. *Expert Systems with Applications*, *39*(1), 1183–1190. https://doi.org/10.1016/J.ESWA.2011.07.124

Bobadilla, J., Ortega, F., Hernando, A., & Gutiérrez, A. (2013). Recommender systems survey. *Knowledge-Based Systems*, *46*, 109–132. https://doi.org/10.1016/j.knosys.2013.03.012

Chen, Y. L., & Cheng, L. C. (2009). Mining maximum consensus sequences from group ranking data. *European Journal of Operational Research*, *198*(1), 241–251. https://doi.org/10.1016/J.EJOR.2008.09.004

Chiang, R. D., Wang, Y. H., & Chu, H. C. (2013). Prediction of members' return visit rates using a time factor. *Electronic Commerce Research and Applications*, *12*(5), 362–371. https://doi.org/10.1016/J.ELERAP.2013.06.002

Choi, K., Yoo, D., Kim, G., & Suh, Y. (2012). A hybrid online-product recommendation system: Combining implicit rating-based collaborative filtering and sequential pattern analysis. *Electronic Commerce Research and Applications*, *11*(4), 309–317. https://doi.org/10.1016/J.ELERAP.2012.02.004

DiClemente, D. F., & Hantula, D. A. (2003). Applied behavioral economics and consumer choice. *Journal of Economic Psychology*, *24*(5), 589–602. https://doi.org/10.1016/S0167-4870(03)00003-5

Duan, Y., Liu, P., & Lu, Y. (2022). MhSa-GRU: combining user's dynamic preferences and items' correlation to augment sequence recommendation. *Journal of Intelligent Information Systems*, 1–24. https://doi.org/10.1007/S10844-022-00754-0/METRICS

Fan, W., Wang, J., Ma, Y., Tang, J., Yin, D., & Li, Q. (2019). Deep social collaborative filtering. *RecSys 2019 - 13th ACM Conference on Recommender Systems*, 305–313. https://doi.org/10.1145/3298689.3347011

Gan, M., Xu, G., & Ma, Y. (2023). A multi-behavior recommendation method exploring the preference differences among various behaviors. *Expert Systems with Applications*, *228*, 120316. https://doi.org/10.1016/J.ESWA.2023.120316

Ganesh, J., Reynolds, K. E., Luckett, M., & Pomirleanu, N. (2010). Online Shopper Motivations, and e-Store Attributes: An Examination of Online Patronage Behavior and Shopper Typologies. *Journal of Retailing*, *86*(1), 106–115. https://doi.org/10.1016/J.JRETAI.2010.01.003

Huang, C. L., & Huang, W. L. (2009). Handling sequential pattern decay: Developing a two-stage collaborative recommender system. *Electronic Commerce Research and Applications*, *8*(3), 117–129. https://doi.org/10.1016/J.ELERAP.2008.10.001

Huang, T. C. K. (2012). Mining the change of customer behavior in fuzzy time-interval sequential patterns. *Applied Soft Computing*, *12*(3), 1068–1086. https://doi.org/10.1016/J.ASOC.2011.11.017

Iwanaga, J., Nishimura, N., Sukegawa, N., & Takano, Y. (2019). Improving collaborative filtering recommendations by estimating user preferences from clickstream data. *Electronic Commerce Research and Applications*, *37*, 100877. https://doi.org/10.1016/J.ELERAP.2019.100877

Kim, Y. S., & Yum, B. J. (2011). Recommender system based on click stream data using association rule mining. *Expert Systems with Applications*, *38*(10), 13320–13327. https://doi.org/10.1016/J.ESWA.2011.04.154

Kim, Y. S., Yum, B. J., Song, J., & Kim, S. M. (2005). Development of a recommender system based on navigational and behavioral patterns of customers in e-commerce sites. *Expert Systems with Applications*, *28*(2), 381–393. https://doi.org/10.1016/J.ESWA.2004.10.017

Koohi, H., & Kiani, K. (2016). User based Collaborative Filtering using fuzzy C-means. *Measurement: Journal of the International Measurement Confederation*, *91*, 134–139. https://doi.org/10.1016/j.measurement.2016.05.058

Koohi, H., & Kiani, K. (2017). A new method to find neighbor users that improves the performance of Collaborative Filtering. *Expert Systems with Applications*, *83*, 30–39. https://doi.org/10.1016/J.ESWA.2017.04.027

Koohi, H., & Kiani, K. (2020). Two new collaborative filtering approaches to solve the sparsity problem. *Cluster Computing 2020 24:2*, *24*(2), 753–765. https://doi.org/10.1007/S10586-020-03155-6

Lee, C.-H., Chen, C.-W., Qi, J., & Širca, N. T. (2021). Impulse Buying Behaviors in Live Streaming Commerce Based on the Stimulus-Organism-Response Framework. *Information 2021, Vol. 12, Page 241*, *12*(6), 241. https://doi.org/10.3390/INFO12060241

Lee, K. C., & Kwon, S. (2008). Online shopping recommendation mechanism and its influence on consumer decisions and behaviors: A causal map approach. *Expert Systems with Applications*, *35*(4), 1567–1574. https://doi.org/10.1016/J.ESWA.2007.08.109

Lei, J., Li, Y., Yang, S., Shi, W., & Wu, Y. (2022). Two-stage sequential recommendation for side information fusion and long-term and short-term preferences modeling. *Journal of Intelligent Information Systems*, *59*(3), 657–677. https://doi.org/10.1007/S10844-022-00723-7/METRICS

Liao, S. hsien, & Chang, H. ko. (2016). A rough set-based association rule approach for a recommendation system for online consumers. *Information Processing & Management*, *52*(6), 1142–1160. https://doi.org/10.1016/J.IPM.2016.05.003

Moe, W. W. (2003). Buying, Searching, or Browsing: Differentiating Between Online Shoppers Using In-Store Navigational Clickstream. *Journal of Consumer Psychology*, *13*(1–2), 29–39. https://doi.org/10.1207/s15327663jcp13-1&2_03

Nazari, Z., Koohi, H. R., & Mousavi, J. (2022). Increasing Performance of Recommender Systems by Combining Deep Learning and Extreme Learning Machine. *Journal of AI and Data Mining*, *0*. https://doi.org/10.22044/JADM.2022.11248.2279

Nishimura, N., Sukegawa, N., Takano, Y., & Iwanaga, J. (2020). Predicting Online Item-choice Behavior: A Shape-restricted Regression Perspective. https://doi.org/10.48550/arxiv.2004.08519

Pirozmand, P., Alrezaamiri, H., Ebrahimnejad, A., & Motameni, H. (2021). A NEW MODEL OF PARALLEL PARTICLE SWARM OPTIMIZATION ALGORITHM FOR SOLVING NUMERICAL PROBLEMS. *Malaysian Journal of Computer Science*, *34*(4), 389–407. https://doi.org/10.22452/MJCS.VOL34NO4.5

Rathipriya, R., & Thangavel, K. (2010). A Fuzzy Co-Clustering approach for Clickstream Data Pattern. *Global Journal of Computer Science and Technology*, *10*. https://doi.org/10.48550/arxiv.1109.6726

Rawat, M., Goyal, N., & Singh, S. (2017). Advancement of recommender system based on clickstream data using gradient boosting and random forest classifiers. *8th International Conference on Computing, Communications and Networking Technologies, ICCCNT 2017*, 1–6. https://doi.org/10.1109/ICCCNT.2017.8204029

Roy, A., & Ludwig, S. A. (2021). Genre based hybrid filtering for movie recommendation engine. *Journal of Intelligent Information Systems*, *56*(3), 485–507. https://doi.org/10.1007/S10844-021-00637-W/METRICS

Schafer, J. ben, Konstan, J. A., & Riedl, J. (2001). E-Commerce Recommendation Applications. *Data Mining and Knowledge Discovery 2001 5:1*, *5*(1), 115–153. https://doi.org/10.1023/A:1009804230409

Su, Q., & Chen, L. (2015). A method for discovering clusters of e-commerce interest patterns using click-stream data. *Electronic Commerce Research and Applications*, *14*(1), 1–13. https://doi.org/10.1016/J.ELERAP.2014.10.002

Wu, J., Li, Y., Shi, L., Yang, L., Niu, X., & Zhang, W. (2022). ReRec: A Divide-and-Conquer Approach to Recommendation Based on Repeat Purchase Behaviors of Users in Community E-Commerce. *Mathematics 2022, Vol. 10, Page 208*, *10*(2), 208. https://doi.org/10.3390/MATH10020208

Wu, R. S., & Chou, P. H. (2011). Customer segmentation of multiple category data in e-commerce using a soft-clustering approach. *Electronic Commerce Research and Applications*, *10*(3), 331–341. https://doi.org/10.1016/J.ELERAP.2010.11.002

Xiao, Y., & Ezeife, C. I. (2018). E-Commerce Product Recommendation Using Historical Purchases and Clickstream Data. *In: Ordonez, C., Bellatreche, L. (eds) Big Data Analytics and Knowledge Discovery. DaWaK 2018. Lecture Notes in Computer Science*, *11031 LNCS*, 70–82. https://doi.org/10.1007/978-3-319-98539-8_6

Yu, H. F., Hsieh, C. J., Si, S., & Dhillon, I. S. (2013). Parallel matrix factorization for recommender systems. *Knowledge and Information Systems 2013 41:3*, *41*(3), 793–819. https://doi.org/10.1007/S10115-013-0682-2

Zhao, W. X., Li, S., He, Y., Wang, L., Wen, J. R., & Li, X. (2015). Exploring demographic information in social media for product recommendation. *Knowledge and Information Systems 2015 49:1*, *49*(1), 61–89. https://doi.org/10.1007/S10115-015-0897-5

Zheng, L., Cui, S., Yue, D., & Zhao, X. (2010). User interest modeling based on browsing behavior. *ICACTE 2010 - 2010 3rd International Conference on Advanced Computer Theory and Engineering, Proceedings*, *5*. https://doi.org/10.1109/ICACTE.2010.5579511